# Electron-electron interactions of the multi-Cooper-pairs in the 1D limit and their role in the formation of global phase coherence in quasi-one-dimensional superconducting nanowire arrays

[1]C. H. Wong*, [1]E. A. Buntov, [1]A. F. Zatsepin, [2]R. Lortz*

[1]Institute of Physics and Technology, Ural Federal University, Yekaterinburg, Russia

[2]Department of Physics, Hong Kong University of Science and Technology, Clear Water Bay, Kowloon, Hong Kong

Nanostructuring of superconducting materials to form dense arrays of thin parallel nanowires with significantly large transverse Josephson coupling has proven to be an effective way to increase the upper critical field of superconducting elements by as much as two orders of magnitude as compared to the corresponding bulk materials and, in addition, may cause considerable enhancements in their critical temperatures. Such materials have been realized in the linear pores of mesoporous substrates or exist intrinsically in the form of various quasi-1D crystalline materials. The transverse coupling between the superconducting nanowires is determined by the size-dependent coherence length $\xi_0$. In order to obtain $\xi_0$ over the Langer-Ambegaokar-McCumber-Halperin (LAMH) theory, extensive experimental fitting parameters have been required over the last 40 years. We propose a novel Monte Carlo algorithm for determining $\xi_0$ of the multi-Cooper pair system in the 1D limit. The concepts of uncertainty principle, Pauli-limit, spin flip mechanism, electrostatic interaction, thermal perturbation and co-rotating of electrons are considered in the model. We use Pb nanowires as an example to monitor the size effect of $\xi_0$ as a result of the modified electron-electron interaction without the need for experimental fitting parameters. We investigate how the coherence length determines the transverse coupling of nanowires in dense arrays. This determines whether or not a global phase-coherent state with zero resistance can be formed in such arrays. Our Monte Carlo results are in very good agreement with experimental data from various types of superconducting nanowire arrays.

## 1. Introduction

Quasi-1D superconductors represent intriguing materials that allow investigating quantum confinement effects, strong phase fluctuations, and critical phenomena, based on the fact that arrays of dense superconducting atomic chains or nanowires represent intrinsic Josephson junction arrays with unique critical phase transition behavior [1,2,3,4,5,6,7]. Of particular interest is the observation that under certain conditions the upper critical fields ($H_{c2}$) can be strongly enhanced with $H_{c2}$ values reaching up to a factor 200 of the value of the corresponding materials in their bulk form, and significant critical temperature ($T_c$) enhancements compared to the bulk state [1,2]. The generation of high magnetic fields based on superconducting solenoids is of great technological interest, for example in the form of high magnetic field electromagnets for nuclear magnetic resonance imaging (MRI) techniques [8] and charged-particle radiotherapy in medicine [9], or for particle accelerators in high energy physics [10]. The nanostructuring superconductors in the form of quasi-1D systems could open new ways to achieve higher magnetic fields with existing superconducting materials, although the nanostructuring of large-scale superconducting quasi-1D nanowire arrays is certainly a great challenge. The $H_{c2}$ enhancement is likely due to the fact that the Fermi surface in a 1D conductor does not allow orbital motion of the charge carriers, thus pushing the orbital limiting field for superconductivity [11] well beyond the Pauli-paramagnetic limit at which the Zeeman energy of the two electrons with spin up and down forming the Cooper pair exceeds the pairing energy [12, 13]. For the $T_c$ enhancements, several reasons have been discussed. The van Hove singularities in the electronic density of states (DOS) in one-dimensional conductors may under certain circumstances lead to very high DOS at the Fermi level, which according to the Bardeen-Cooper-Schrieffer (BCS) theory for superconductivity, is one of the most important ingredients for a high $T_c$ [14,15]. In addition, quantum confinement may play a favorable role in improving the

ch.kh.vong@urfu.ru
lortz@ust.hk

superconducting parameters [16], as well as phonon softening at the surface of the nanowires [1]. Quasi-1D superconductors have also stimulated the development of the theory of 'superstripes' [17,18,19].

A variation of the wavefunction in 1D superconducting nanowires with a transverse dimension smaller than the superconducting coherence length is energetically forbidden over the cross section of the nanowire, so that the wave function only varies as a function of the chain axis [20]. Under these circumstances, the occurrence of phase-slip processes suppresses the global phase coherence of a purely 1D superconductor [20]. The phase-slips along the nanowire cause a finite electrical resistance below $T_c$, although Cooper pairs are already well formed [20]. However, this limitation can be overcome by arranging the nanowires to form dense parallel arrays [1,5,7,21] or even random networks [2], which is also realized by superconducting atomic chains in some intrinsic quasi-1D superconductors including $Sc_3CoC_4$ [3] and $Tl_2Mo_6Se_6$ [4]. In order to completely suppress the phase fluctuations and create a phase-coherent three-dimensional (3D) superconducting bulk state in the array, the superconducting order parameters in the individual 1D superconducting elements must undergo a macroscopic phase ordering transition that can only be triggered when there is a significant coupling. This can be achieved by Josephson coupling between parallel nanowires embedded in an insulating host, such as $AlPO_4$-5 (AFI) zeolite [4,7] or mesoporous SBA-15 silica [1,21], or in the form of the proximity effect, such as realized in $Tl_2Mo_6Se_6$ [4]. This phase ordering transition triggers a dimensional crossover in the array from a 1D fluctuating superconducting state with finite resistance at high temperatures to a zero-resistance 3D bulk phase-coherent superconducting state in the low temperature regime. It has been shown that this transition falls into the same universality class as the famous Berezinski-Kosterlitz-Thouless (BKT) transition [22,23,24] in 2D superconducting systems [6], where a vortex-unbinding transition occurs through vortex excitations in the phases of the order parameters in groups of adjacent nanowires [3,4]. The Josephson phase current, which establishes the BKT-type transverse coupling, is very weak when the tunneling barrier is thick or the coherence length is short. However, the superconducting coherence length typically depends on size. For example, the coherence length of Pb is shortened from 83nm to 50nm when the geometry is transformed from 3D to 1D [25]. Although the Langer-Ambegaokar-McCumber-Halperin (LAMH) theory is well accepted for the description of 1D superconductivity, the direct calculation of the coherence length requires experimental fitting parameters [20,26]. This stimulated us to develop a theoretical approach that can directly investigate the role of the coherence length in the 1D limit and its effects on the BKT-like phase-ordering transition, which triggers the dimensional crossover in quasi-1D superconductors.

## 2. Computational methods

A Cooper pair in a superconducting spin-singlet state represents a loosely bound electron pair with anti-parallel spins in which the electrons travel in opposite directions at the same magnitude of velocity [14]. The Hamiltonian $\Delta H$ of the Cooper pair relative to the ground state can be written as follows:

$$\Delta H = \sum \left[ \left| \frac{q_1 q_2}{4\pi\varepsilon r_{12}} - \frac{q_1 q_2}{4\pi\varepsilon r_c} \right| + \frac{1}{2} m \left( \left| v_1^2 - v_G^2 \right| + \left| v_2^2 - v_G^2 \right| \right) + \frac{1}{2}(2m)\left(V_{CM}^2 - V_{CMG}^2\right) + \frac{1}{2}\mu H_c^2 (\cos\theta + 1)\delta\zeta_{12} \right]$$

where $q_1 = q_2 = -1.6x10^{-19}$ Coulomb is the electrostatic charges of the electrons 1 & 2, respectively. $r_c$ is the bulk coherence length at 0K and $r_{12}$ is the separation of electrons at any finite temperature. The velocity of individual electron are defined by $v_1$ and $v_2$. The minimum speed of a single electron due to the uncertainty principle at 0K is expressed as $v_G \sim \frac{\hbar}{2mr_c}$ [27]. $V_{CM}$ is the velocity of the center of mass at finite temperatures, while $V_{CMG} = 0$ is the ideal common velocity at 0K. The mass of electron $m$ is given as 9.11x10$^{-31}$ kg. The magnetic permeability $\mu$ and electric permittivity $\varepsilon$ are 1.25x10$^{-6}$ and 8.85x10$^{-12}$, respectively. $H_c \sim 1.84 T_c$ is the critical field due to the Pauli limit [28]. $\theta$ is the relative angle of the spins between the electrons and $\delta\zeta_{12}$ is the coherence volume. The dynamics of the electrons are simulated in the asymmetric 2D XY plane. The known parameters for bulk Pb will be imported in the Hamiltonian. In a BCS superconductor the separation of electrons satisfies $r = r_c \left(1 - \frac{T}{T_c}\right)^{-1}$ at any finite temperature $T$, where $T_c$ and $r_c$ are 7.2K and 83nm respectively [14]. The motion of the electron in each step must be less than the coherence length at 0K to ensure that the iterative path is physical. The resistivity of bulk Pb at ~10K is 1.35x10$^{-10}$ $\Omega$m [29]. The scattering time $\tau$ is then converted from the resistivity via the Drude model. Therefore, the smallest iterative path is defined as $\delta\ell = v_G \tau \left(1 - \frac{T}{T_c}\right)^{-1}$.

The initial condition is addressed so that the angle $\theta$ is set to 180 degrees, $V_{CM}$ is zero and $v_1 = v_2 = v_G$, initially. The initial $r_{12}$ equals to $0.5 r_c R_{ran}$, where $0 \leq R_{ran} \leq 1$. The verified consequence of choosing some pre-factors that differ strongly from 1 will increase the computational costs without having a noticeable effect on the final relaxed energy, and so we use a pre-factor of 0.5 as a good trial value.

The Monte Carlo algorithm begins by randomly selecting the electron and then comparing the energies between the 4 nearest neighbors in the inner part. On the other hand, only three nearest electrons can be found along the edges. This electron selects one of the nearest neighbors. The Monte Carlo metropolis algorithm recommends relaxing the energies from more positive to less positive states, and so the first empirical estimate is to consider the energy between the selected electron and the most energetically unfavorable neighbor. Based on the iterative path, the selected electron attempts to move along the preferred axis. The sign of the preferred axis is followed by another random number. If the random number is greater than or equal to 0.5, the sign of the preferable axis is positive. Otherwise, the preferred direction becomes opposite [30]. In our simulation, four rotational states, i.e. 0, 90, 180, 270 degree, are registered. The spin is allowed to flip to the adjacent angle during each Monte Carlo step. For example, if the initial spin points to 90 degrees, the possible trial state is either 0 or 180 degrees at the same probability of attempting at each Monte Carlo step. Now the trial $r_{12}, v_1, v_2, V_{CM}, \theta, \delta\zeta_{12}$ values are known. If the new trial status reduces the total energy, the trial status is accepted to lock the electron pairs. Otherwise, they will return to the original status. The system is relaxed to equilibrium and meanwhile the electron pairs can change their partners as a function of Monte Carlo steps. At equilibrium, the singlet orientation angle of the Cooper pairs is assigned to be the spin state of one of the electrons within the pairs. We have noted that the electrons in the Cooper pairs must have opposite spin and in the meantime keep relaxing the system by updating or rejecting the new $r_{12}, v_1, v_2, V_{CM}$ and $\delta\zeta_{12}$ values simultaneously. The Boltzmann factor is monitoring the thermal excitation followed by the metropolis approach at finite temperatures [31]. In this simulation, the size of array will be decreased from 64 x 64 to 64 x 6, gradually, in order to study how the Cooper pairs behave when approaching the 1D limit. After deducing the coherence length in individual nanowires in the purely 1D limit, the strength of the transverse Josephson coupling, which triggers the BKT-like phase-ordering transition, will be compared in various types of nanowire arrays. The maximum Josephson current is expressed by $J_{max} = \frac{4q\hbar\alpha}{m} \frac{n_e}{e^{2\alpha a} - e^{-2\alpha a}}$ where $\alpha = \frac{\sqrt{2mU}}{\hbar}$, $n_e$ is electron concentration, $a$ is the lateral spacing between the nanowires and $U$ is the tunneling barrier [15]. We borrow the concept of a characteristic BKT temperature ($T_{BKT}$) below which vortex-antivortex pairs in the phases of the individual order parameters of small groups of adjacent nanowires form bound pairs [22,23,24] and thus cause the formation of the global phase coherent state in the entire array [1]. The Josephson energy $E_{max}$ and $T_{BKT}$ are directly proportional to the strength of the transverse Josephson coupling and are determined by $E_{max} = \frac{J_{max}\hbar}{2q}$ and $T_{BKT} \sim \frac{\pi E_{max}}{1.12 k_B}$, respectively [4].

## 3. Results and discussion

The normalized energies of the Pb nanowire with an aspect ratio of 0.18 are plotted in Fig. 1. After all the simulated electrons have been adjusted to their initial conditions, we find that the relaxed energies are stabilized beyond 250000 Monte Carlo steps. The energies are at large positive values at the beginning of the simulation because the electrons are compressed, where the $r_{12}$ becomes the dominant factor in the Hamiltonian. After unlashing of the electrons, the energies decrease and become less positive, because of $r_{12} \sim r_c$. A faster decrease in energy is observed at 3K, and the energy fluctuations in equilibrium are also stronger at 3K due to the stronger Boltzmann excitation, which increases the kinetic energy of the electrons [32]. The energy in equilibrium at 3K is more positive than at 1K. The reasons for this are, on one hand, that the stronger thermal excitation increases the distance between the electrons from the ground state value, which corresponds to the 1st term in the Hamiltonian. On the other hand, thermal energy causes the velocity of electrons to deviate more from $v_G$ and thus presumably the 2nd and the 3rd terms in the Hamiltonian increase.

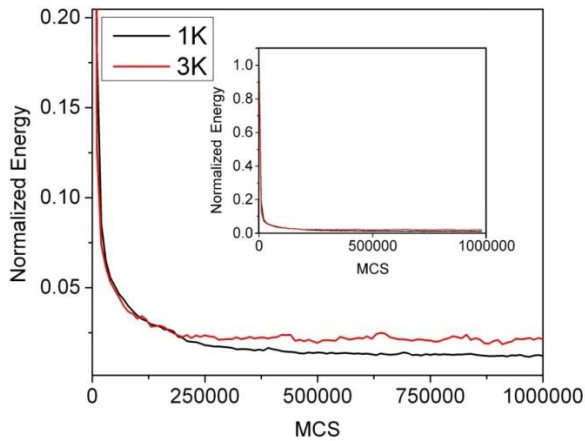

Fig 1: Normalized energy of the Pb nanowire as a function of Monte Carlo steps (MCS) at different temperatures. The inset shows the full energy scans. The array size is 64 x 12.

Fig. 2 shows that the coherence length is shortened from 79nm to 53nm, which agrees well with the experimentally observed shortened coherence length of 50nm in Pb nanowires of similar dimensionality [25]. The size dependence of the coherence length is more apparent when the aspect ratio is less than 0.4, since the surface to volume ratio is inversely proportional to the reciprocal of the radius. The asymmetric electron-electron interactions due to the geometric boundaries become significant at low aspect ratios, and therefore the superconducting fluctuations in thinner nanowires are stronger [20,26]. This makes the $\Delta H$ much larger via readjusting the $r_{12}$, $V_{CM}$, $v_1$, $v_2$ or $\theta$. The thinner nanowire therefore restricts the electrons more and causes shorter coherence lengths [20]. In the following we will use Pb nanowires with an aspect ratio of 0.09 as an example. The fluctuations in energy in equilibrium contributed by the spin term are less than 1%. However, ~51% of the fluctuations are due to the electrostatic term, while the individual kinetic energies and the common kinetic energy of the Cooper pairs contribute ~26% and ~23%, respectively. According to the BCS theory, $T_c$ is inversely proportional to the coherence length [14].

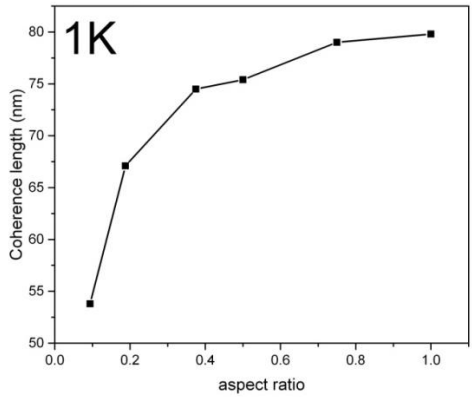

Fig 2: Reduction in the coherence length of Pb during a simulated crossover from a 2D film to a 1D nanowire.

Fig.3 shows that the superconducting transition temperature increases with decreasing aspect ratio. Fig. 4 shows how the coherence length behaves upon increasing temperature. The coherence length of the Pb nanowire with the aspect ratio of 0.18 is increased from 67nm to 73nm as the temperature is increased from 1K to 5K. This due to the increase of the Boltzmann factor at higher temperatures [14,15], and presumably the mean free path of the electrons increases. Apart from that, a dramatic increase in the coherence length beyond 6K is observed, as the correlation between the electrons is almost lost as the critical temperature approaches, and finally the attractive potential between the electrons vanishes compared to the increasing strength of pair breaking thermal fluctuations in the critical regime [14,20]. A similar situation is identified in the common velocity of the electron pairs, as shown in Fig. 5. As the thermal energy is larger, the electrons tend to give up the symmetry of the Cooper pairs by attaining a common velocity. However, the Cooper pairs are practically at rest because the common velocities are lower than 70m/s, which according to the uncertainty principle is almost less than 10% of the minimum velocity range. As a result, we confirm that our Monte Carlo simulation does not destroy the symmetric co-rotation of the paired electrons regardless of the temperature.

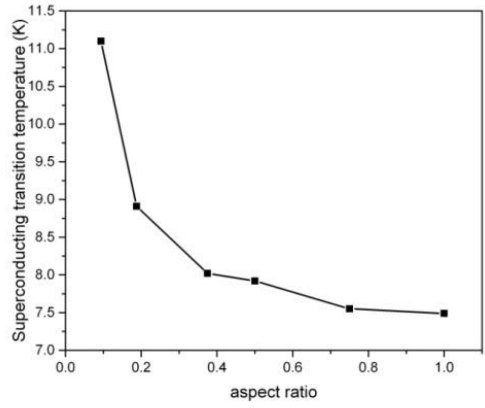

Fig 3: Superconducting transition temperature $T_c$ of Pb as a function of aspect ratio of the nanowire. $T_c$ increases when approaching the 1D limit.

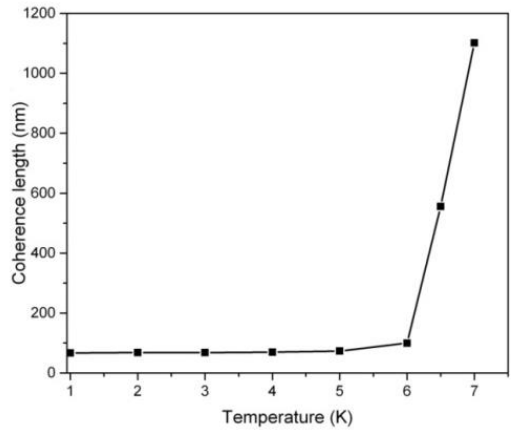

Fig 4: Thermal effect on the coherence length of the Pb nanowire for an aspect ratio of 0.18.

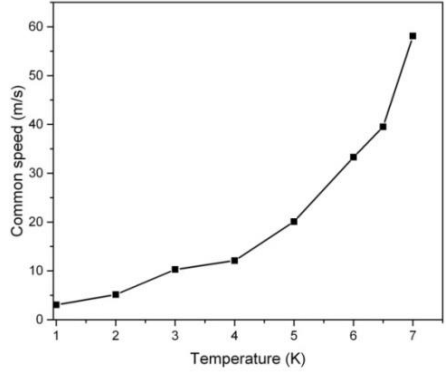

Fig 5: Center-of-mass velocity of the Cooper pairs in a Pb nanowire as a function of temperature for an aspect ratio of 0.18.

We calculate $T_{BKT}$ of various superconducting nanowires arrays embedded in a mesoporous silica substrate, which has been experimentally realized in Ref. 1 and 21 for Pb and NbN. The experimental data for Pb nanowires, which are laterally separated by 5nm, show that $T_{BKT}$ is about 5.5K and the onset of the superconducting transition temperature $T_c$ is at 11K [1]. Considering an insulating substrate with a band gap of ~1 eV, the theoretical $T_{BKT}$ is 4.7K in Pb nanowire arrays, where an electron concentration of 3.2 x $10^{28}$ m$^{-3}$ was considered. By using the same approach, the maximum $T_{BKT}$ of various types of superconducting nanowires embedded in a silica substrate is listed in Table 1, assuming that the coherence length is longer than the insulating separation width [15]. Although the onset $T_c$ of the NbN nanowires array is the largest, its low electron concentration is one of the main reasons that $T_{BKT}$ is as low as 0.007K, which agrees with recent experimental observations [21]. In contrast, the $T_{BKT}$ of Sn nanowires almost matches the experimentally observed $T_c$ value in networks of freestanding Sn nanowires [2]. Compared to the experimental Sn data, the theoretical $T_{BKT}$ ~5K is higher than the measured value [2]. This discrepancy is likely due to the irregular random alignment in the nanowire network in Ref. 2. The regularity of the nanowires thus plays an important role in the strength of the BKT-like transition [14,33]. A recent study of the BKT coupling in two coaxial quasi-1D superconducting cylindrical surfaces showed that $T_{BKT}$ depends on the separation between the superconducting nanowires [33]. An irregularity in the alignment of nanowires weakens the strength of the mean Josephson interaction so that $T_{BKT}$ is reduced.

Table 1: The maximum BKT-like transition temperatures of the superconducting nanowires arrays.

| Superconducting nanowires array | $T_{BKT}$ /K |
|---|---|
| Pb | 4.7 |
| NbN | 0.007 |
| Sn | 5.1 |

## 4. Conclusion

By addressing the parameters of the uncertainty principle, Pauli-limit, spin-flip mechanisms, electrostatic interactions, thermal perturbation and the co-rotation of electron pairs with our newly developed Monte Carlo algorithm, we have addressed one of the unsolved issues in the Langer-Ambegaokar-McCumber-Halperin (LAMH) theory. The estimation of the coherence length of the 1D superconductor no longer requires experimental fitting parameters. The prediction of the 1D coherence length is very important to interpret the maximum BKT-like phase ordering transition temperatures that occur in dense superconducting nanowire arrays and controls a transition from a 1D fluctuating superconductivity at high temperatures to a 3D phase coherent bulk superconducting state with zero resistance in the low-temperature regime over the transverse Josephson coupling.